\documentclass{sig-alternate}

\usepackage{latexsym}
\usepackage{graphics}
\usepackage{color}
\usepackage{amssymb,amsmath,amsfonts}
\usepackage{wrapfig,epsfig}
\usepackage[center]{caption}
\usepackage{subfigure}
\usepackage{color}
\usepackage{algorithm}
\usepackage{algpseudocode}
\usepackage{url}
\usepackage{paralist}
\usepackage{enumitem}
\usepackage{adjustbox}

\newcommand{\pj}{\mbox{{${\textbf{\#\!\;PJ}}$}}}
\newcommand{\lp}{\mbox{{${\textbf{\#\!\;LP}}$}}}
\newcommand{\jo}{\mbox{{${\textbf{\#\!\;JO}}$}}}
\newcommand{\pp}{\mbox{{${\textbf{\#\!\;PP}}$}}}

\newcommand{\TPFQ}[2]{\mbox{{${\bf PF} (#1, #2)$}}}

\newcommand{\TOF}[1]{\mbox{{${\bf OF} (#1)$}}}

\newcommand{\run}[2]{\mbox{{${\tt r} (#1, #2)$}}}

\def\compactify{\itemsep=0pt \topsep=0pt \partopsep=0pt \parsep=0pt}
\let\latexusecounter=\usecounter

\begin{document}
\begin{sloppypar}

\title{OptMark: A Toolkit for Benchmarking Query Optimizers}

\numberofauthors{1}
\author{\alignauthor
Zhan Li$^*$, Olga Papaemmanouil$^*$ and Mitch Cherniack$^*$\\
       \affaddr{$^*$ Brandeis University, Waltham, MA, USA}\\
       \email{$^*$\{zhanli,olga, mfc\}@cs.brandeis.edu}
}

\maketitle
\begin{abstract}
Query optimizers have long been considered as among the most complex components of a database engine, while the assessment of an optimizer's quality remains a challenging task. 
Indeed, existing performance benchmarks for database engines (like TPC benchmarks) produce a performance assessment of the  query runtime system rather than its query optimizer. To address this challenge, this paper introduces \emph{OptMark}, a toolkit for evaluating the quality of a  query optimizer.  OptMark is designed to offer  a number of desirable properties. First, it decouples the quality  of an  optimizer from the quality of its underlying execution engine. Second it evaluates independently both  the effectiveness of an optimizer (i.e., quality of the chosen plans) and its efficiency (i.e., optimization time). OptMark includes also a generic benchmarking toolkit that is  minimum invasive to the DBMS that wishes to use it. Any DBMS can provide a system-specific implementation of a simple API that allows OptMark to run and generate benchmark scores for the specific system.  This paper discusses the metrics we propose for evaluating an optimizer's quality, the benchmark's design and  the toolkit's API and functionality. We have implemented OptMark on the open-source MySQL engine as well as two commercial database systems. Using these implementations we are able to  assess the quality of the optimizers on these three systems based on the TPC-DS benchmark queries.

\end{abstract}

\section{Introduction}\label{s_introduction}

Query optimizers have been recognized as among the most complex components of a DBMS.
Among the myriad of optimizer design choices are whether they are top-down or {bottom-up \cite{dehaan}}, how (or if) they
constrain the search space of possible {plans \cite{enumeration-fender}}, or whether or not plans are modified {dynamically \cite{adaptive}}.
Despite the wide variety in approaches to query optimization, the assessment of an optimizer's {\em quality} remains highly subjective.
Indeed, DBMS performance benchmarks (e.g., the TPC benchmarks) conflate
query optimization and query execution, producing a performance assessment that
reflects upon the DBMS' query runtime system as much as (and arguably more than) its query optimizer.

Undoubtedly, a major reason that no query optimizer benchmark exists is because such a benchmark
is extremely difficult to design and implement\cite{wass-blog11}.  We have identified the following
three key challenges to the design of an effective optimizer benchmark:

\begin{itemize}[leftmargin=*]
\item {\bf Dual Assessment Measures:}  Optimizers should be evaluated for both their {\em effectiveness} and {\em efficiency} in generating
plans for a given suite of queries:

\begin{itemize}[leftmargin=*]
\item [] {\it Effectiveness:}  An optimizer benchmark must measure the quality of plans generated for queries in a given query suite.
But in comparing the optimizers of two different DBMSs, it is insufficient to compare the execution times of plans chosen by the optimizers
for the same query, even if both DBMSs are implemented on the same platform.  To illustrate, a plan generated by the MySQL optimizer~\cite{mysql}
may fare poorly compared to that of commercial DBMS for a join-heavy query because joins in MySQL queries can only be executed as
{\em nested-loop joins}.  This does not necessarily reflect the effectiveness of the MySQL optimizer which might consistently
generate the highest-performing plans possible that exclusively use nested-loop joins.  Thus, optimizer effectiveness must
be assessed evaluating generated plans relative to other plans the DBMS' query execution engine is capable of running.

\item [] {\it Efficiency:}  An optimizer benchmark must also measure the resources (i.e., time and space) required by an optimizer
to generate plans.  A given optimizer could be very effective if it combines {\em exhaustive search} (thus considering
all possible plans for a given query) with an extremely accurate cost model. But in practice, it is infeasible to exhaustively
consider all possible plans, especially for the most complex and expensive queries (e.g., queries involving large numbers of tables)
where optimization is needed most.\footnote{This is the reason that most optimizers have timeout settings that allow optimization
to be curtailed prior to consideration of all plans.}  Thus, efficiency is a measure of how well a query optimizer can scale to
process the most complex of queries.
\end{itemize}

\item  {\bf Benchmark Generality:}  An optimizer benchmark should be runnable over any DBMS regardless of the (hardware and OS) platforms
over which it runs, and benchmark scores for different optimizers should be comparable even when run on differing platforms.  This implies that:

\begin{itemize}
\item benchmark code should be configurable to any DBMS (but ideally requiring minimal DBMS-specific code to do so), and
\item time-based metrics of effectiveness (runtime of generated plan for given query) and efficiency (time spent optimizing given query)
should be avoided as they are incomparable for systems running on different platforms
\end{itemize}

\item  {\bf Isolated Assessment:}  A DBMS optimizer's performance must be decoupled from that of the DBMS' query execution engine.
Thus, end-to-end benchmarks such as the TPC-H and TPC-DS benchmarks \cite{tpcds} are not good query optimizer benchmarks because they report
query execution times which depend not only on the plan chosen by an optimizer but (even more so) on the capabilities of the DBMS
query execution engine.
\end{itemize}

\noindent In this paper, we introduce {\em Opt-Mark}: a Query [Opt]imizer Bench[Mark] with the following key features:

\begin{enumerate}
\item {\em Effectiveness metrics} that assess the performance of optimizer-chosen plans relative to other plans that can be run by the same DBMS.

\item {\em Efficiency metrics} that are not based on optimization timing but instead on the size of the search space that an optimizer considers (thus measuring both space and time).

\item {\em A Toolkit} consisting of generic benchmark code together with a concise API that must be implemented for a benchmarked DBMS in a system-specific way.
\end{enumerate}

This paper is structured as follows. We present the benchmark design, including effectiveness and efficiency measures in Section~\ref{s_design},  and
describe the toolkit code, including the API requiring DBMS-specific implementation in Section~\ref{s_toolkit}.  We present benchmark results for 3 systems:  MySQL and two well-known commercial DBMSs (which we
refer to as Systems X and Y respectively) in Section~\ref{s_results}.  We describe other work related
to query optimizer benchmarks in Section~\ref{s_related} and conclude with our final remarks in Section~\ref{s:conclusions}.

\section{Benchmark Design}\label{s_design}

In this section we present the design of OptMark, describe its {\em effectiveness} and {\em efficiency} measures and
techniques used for determining them.

\subsection{Optimizer Effectiveness}\label{s:effectiveness}

The {\em effectiveness} of a DBMS' optimizer reflects the quality of the plans it generates. The main challenge here is isolating the effectiveness of the optimizer from the underlying DBMS's query execution engine. We discuss this challenge and  we introduce the  effectiveness metrics used in OptMark. 

{\bf Isolated Assessment} We argue that effectiveness should be evaluated  relatively to the capabilities of the underlying query (runtime) engine so as to decouple the effectiveness assessment of this component.  In other words, an optimizer should not be penalized for not considering query operations (i.e., join algorithms, access methods, etc) that are not supported by the runtime engine of the DBMS.  

The necessity of decoupling the optimizer from the query engine is illustrated in the case of MySQL \cite{mysql} whose query engine supports {\em nested-loop joins} (NLJ) as the sole means of evaluating joins.  Especially for join-heavy queries, MySQL will frequently be outperformed by DBMSs that also support other join operations such as {\em sort-merge} and {\em hash} joins.  However, the MySQL optimizer should be considered effective if it consistently identifies the best NLJ plan for a given query, even though the MySQL  query  engine is less effective than those that can perform other types of joins. 

Driven from the above discussion, we introduce the concept of a \emph{relative optimal plan} of a  query in a given DBMS. The relative optimal plan refers to the best plan the  DBMS can run for that query. This plan might be different across different DBMSs. 

{\bf Effectiveness Metrics} OptMark measures optimizer effectiveness using two metrics that both compare the plans that an optimizer chose for a given query suite $Q$ with the plans it could have chosen for the same query engine:

\begin{enumerate}
\item {\em Performance Factor:} For any query $q \in Q$ and optimizer $O_D$ for DBMS $D$, the {\em Performance Factor} of $O_D$ relative to $q$, $\TPFQ{O_D}{q}$, measures the proportion of plans in the search space that are worse than the optimizer-chosen plan, which is defined as 
\begin{equation}\label{eq:PF}
 \TPFQ{O_D}{q}=\frac{\left\vert \{ p | p \in P_D(q), \run{D}{p} \ge \run{D}{O_D(q)}\} \right\vert}{\left\vert{P_D(q)}\right\vert}
 \end{equation}
such that $O_D(q)$ is the plan $O_D$ generates for $q$, $P_D(q)$ is the set of all plans that could be executed by $D$ to evaluate $q$, $\run{D}{p}$ is the measured runtime of plan $p$ over $D$ and \run{D}{O_D(q)} is the runtime of the plan $O_D$ generates for $q$ over $D$. We note here that the timing of $\run{D}{p}$ is subject to numerous environment conditions (e.g., empty DB and OS buffers, no contention) and one should either control these factors (e.g., by executing the queries in isolation and using a  cold cache every time).  
Thus, the best possible score for this metric is 1 (indicating that an optimizer chooses the optimal plan for the query $q$) and the closer the score is to 0, the poorer is the optimizer-chosen plan.

\item {\em Optimality Frequency:}  Optimizer $O_D$ finds a relative optimal plan for query $q\in Q$ if $\TPFQ{O_D}{q} = 1$. Thus, the  optimality frequency of $O_D$, $\TOF{O_D, Q}$, is the percentage of queries in the query set $Q$ for which $O_D$ chose the relative optimal plan.


\end{enumerate}

{\bf Effectiveness Metrics Discussion}  The two metrics described above can be leveraged to provide insight on (a) the quality of the optimizer-chosen plan, (b) the quality of the cost model and (c) the quality of the plan enumeration process of a given optimizer.  The performance factor  reflects the quality of the optimizer-chosen plan compared to other plans the DBMS is capable of running relative to a given query. A performance factor of 1 indicates that the optimizer-chosen plan is better than \emph{all} plans while the lower the performance factor the more plans are better than the chosen one.

One can also examine the plans that did better than the optimizer-chosen plan and get a sense of the quality of cost model and the quality of plan enumeration approach used by an optimizer.  Specifically, the quality of the cost model can be measured by determining how many of the plans that did better than the optimizer-chosen plan were considered by the optimizer. The quality of the plan enumeration can be measured by the number of plans that did better than the optimizer-chosen plan and were not considered by the optimizer. For the plans that were considered by the optimizer, the optimizer did not choose them because of  an  inaccurate cost model. For the plans that were not considered by the optimizer, the optimizer did not include them in its search space due to poor plan enumeration.

Here, we want to emphasize that in the design of the optimizer, there is always a trade off between the quality of the plan and the resources the optimizer uses to find the plan. While ideally the goal of an optimizer is to discover the best plan, in practice optimizers aim to find a good enough plan within limited time available for query optimization. 
Hence, our effectiveness evaluation should  factor along the efficiency of an optimizer. Here, there is trade-off to be noticed. An optimal plan enumeration would consider all possible plans, but that would lead to an inefficient optimizer. On the other hand, an optimizer that considers a few plans is very efficient but if none of these plans are relative optimal or even good enough then the plan enumeration is of low quality.

\subsubsection{Plan Space Generation}\label{ss:cpg}

Evaluating the effectiveness metrics requires the enumeration and execution of \emph{all} possible plans the optimizer $O_D$ could execute for a given query $q$, $P_D(O_D, q)$. This process is unfeasible, especially for  complex queries due to the exponential number of queries to be executed.   Alternatively, one could  collect those plans that optimizer considered (i.e., costed) in the process of choosing a plan. We can then compare these plans with the optimizer-chosen plan and calculate the above effectiveness metrics. However, this approach ignores flaws that might exist in the optimizer's enumeration strategy that might have resulted in an optimizer not considering a plan that it should have. For example an optimizer that only considers poor plans but costs them correctly would be considered effective (i.e., with have  performance factor of 1)   despite choosing poor plans.


OptMark takes an alternative approach and generates a set of comparable plans for a given query by generating a random sample set of candidate execution plans that may or may not have been considered by the optimizer. Our approach assesses the quality of the optimizer-chosen plan relative to the plans that the runtime engine is capable of executing. The challenge here is to estimate the proportion of plans that perform worse than the optimizer's chosen plan (i.e., the performance factor) without having any prior knowledge of the performance distribution of the candidate plans. In the following section we describe how to identify the size of the sample set required to estimate the performance factor with a given confidence and precision. We then proceed to describe how we generate these sample plans.




\subsubsection{Sample  Size}\label{s:samplesize} 

To calculate the sample size required to estimate (with a specified level of confidence and precision) the performance factor  of $O_D$ relative to a query $q$, $\TPFQ{O_D}{q}$, we use the formula introduced in~\cite{random-sampling}: 
\begin{equation}\label{eq:samplesnum}
 n~=~\frac{Z^2p(1-p)}{e^2}
\end{equation}
such that, $e$ is the desired level of precision (aka sampling error), $Z$ is the value from the standard normal distribution that corresponds to the desired confidence level (e.g., $Z=1.96$ for a a confidence interval of 95\%) and $p$ is an estimate of the proportion of plans that will be worse than the optimizer's plan or  0.5 when this estimate is unknown. Equation~\ref{eq:samplesnum}  assumes that our candidate plan set size (i.e., population size) is large compared with the sample size while for smaller populations a modified formula can be used that reduces slightly the required sample size. 

Based on Equation~\ref{eq:samplesnum} if the user desires to estimate the performance factor  of a query with a 95\% confidence level and 5\% precision, it needs to generate at least 385 random sample plans. It follows that if we randomly generate 385 plans and conclude that 80\% of the plans perform worse the optimizer's chosen plan, we can conclude with 95\% confidence that the optimizer's plan is better than 75\% - 85\% possible plans, i.e., the performance factor is 0.75 - 0.85.



\subsubsection{Random Plan Generation}\label{s:joinordering}

 Given a number of random plans to generate, OptMark generates these plans by exploring the three main features that characterize query  plans: the {\em join ordering}, the {\em physical join algorithm} used for each join and {\em table access methods}.  Our process first produces a random join ordering for a given query and then expands this  plan with randomly selected physical join operations and  access methods. Next we describe this process starting with the random join ordering generation for a given query.

\begin{figure}[t]
\centering
\includegraphics[width=2.5in,angle=0]{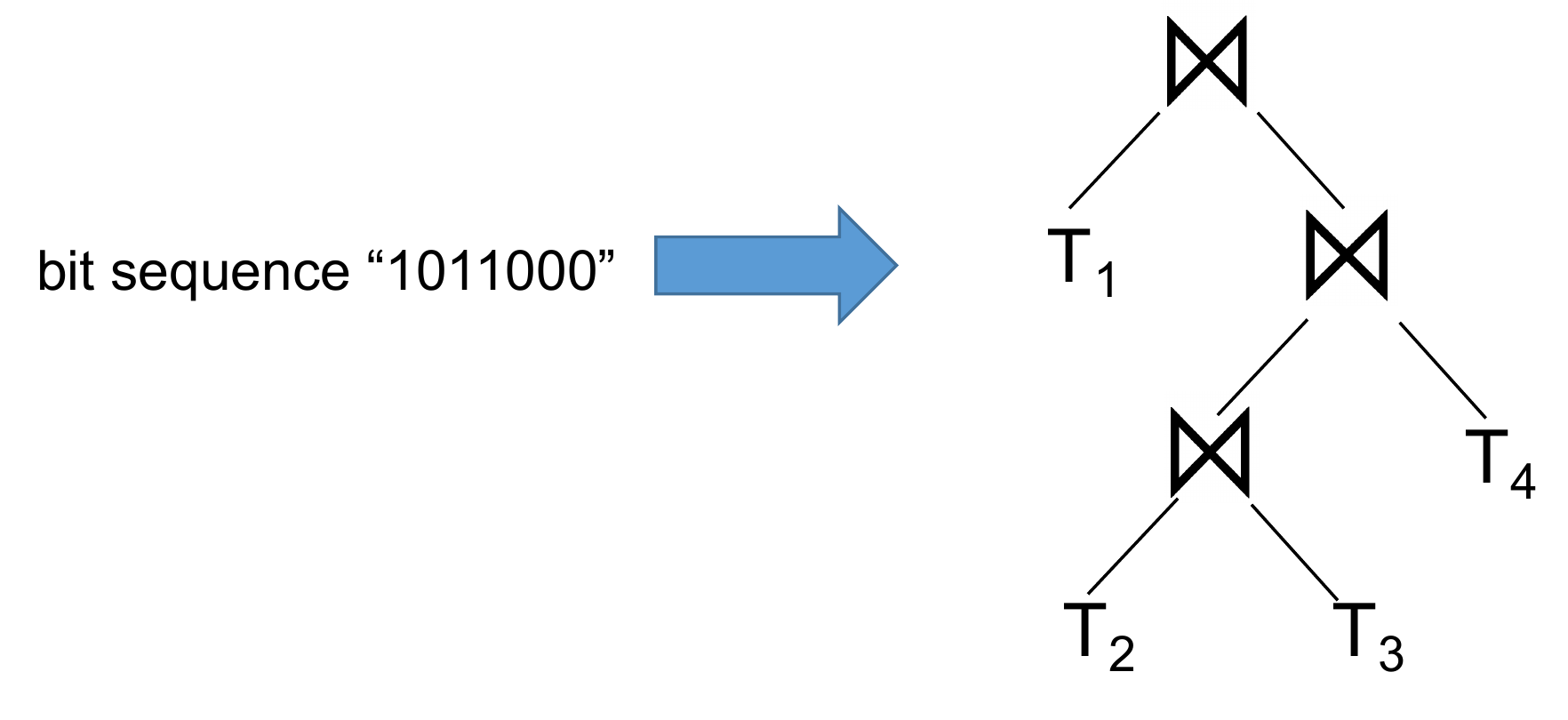}
\caption{``Decode'' a bit sequence to a join tree}
\label{f:join-tree}
\end{figure}

{\bf Random Join Orderings} Traditionally, join orderings are represented as binary trees (aka join trees) where internal nodes represent join operations and the leaf nodes represent tables. Our join ordering generation process generates an unbiased random binary join tree and then generates a random sequence of tables to populate the leaves of the join tree. Every join ordering of a query  joining $n$ tables can be encoded as a pair $(s,p)$ such that:

\begin{itemize}
\item $s$ is a bit sequence that represents the preorder traversal of the ordering's binary tree such that each successive bit denotes the next node of the tree visited in the traversal and is a `1' if that node is an internal node (join)  and a `0' if that node is a leaf node (table). Hence, the bit sequence has length $(2n-1)$ consisting of $(n-1)$ `1's and $n$ `0's. For example ``1011000'' is the encoding for the join tree shown in Figure~\ref{f:join-tree}.


\item $p$ represents the sequence of tables to populate the leaf nodes in the binary tree. Specifically, $p$ is some permutation of the tables in the query. The permutation sequence then specifies the leaves of the join tree from left to right. For example, if the query has tables A, B, C and D, then the permutation ``CBDA" would set $T_1$ = C, $T_2$ = B, $T_3$ = D and $T_4$ = A in the join tree shown in Figure~\ref{f:join-tree}.

\end{itemize}

%
%

%
%
%

OptMark generates random join orderings by generating random encodings of join trees $(s,p)$.  The random generation of the bit sequence $s$ involves flipping a biased coin for each bit in the sequence (going from left-to-right) to decide if the ``next" bit should be a `0' or an `1'~\cite{random-tree}. Because the last bit in the bit sequence must be a `0', given that the last visited node in a preorder traversal must be a leaf node, we use the algorithm in \cite{random-tree} to generate $2(n-1)$ bits at random and add a `0' at the end. 
To ensure that every tree is enumerated with equal probability, one must use a biased coin when deciding between a `1' and a `0' and the degree of bias depends on what has been generated thus far. For each bit in the sequence $s$, we determine the probability that the bit should be filled with a `0', which is expressed as $P(r, k)$ such that $r$ is the number of `1's in the bit sequence thus far minus the number of `0's in the bit sequence thus far and $k$ is the number if bits that have yet to be assigned. The formula to calculate $P(r, k)$ as written in \cite{random-tree} is:
\begin{equation}
P(r, k) = \frac{r  (k+r+2)}{2  k  (r+1)}
\end{equation} 
Given a query that joins $n$ tables, OptMark uses the above formula to estimate the probability of each bit in a $2(n-1)$ bit sequence and create a random sequence $s$. It then generates a random permutation $p$ of the tables in the query and outputs a random join ordering $(s,p)$.

Given the join ordering, we then replace all join operators with randomly selected physical join operators to generate a physical plan to include in the sample.  Specifically, for each join  node in the join tree, if its inputs have no corresponding join predicate in the query, we force a cross join. 
 Otherwise, we randomly select one of the   physical join operators supported by the execution engine. We then add random access methods for each input table by randomly selecting an applicable index or, if one does not exist,  a sequential scan as the table access method. The above process is repeated until we generate as many plans as specified by the sample size  determined as described in Section~\ref{s:samplesize}.

\subsection{Optimizer Efficiency}\label{s:efficiency}

The {\em efficiency} of a DBMS' optimizer reflects the resource requirements (i.e., time and memory) necessary for the optimizer to
choose a query plan.  In theory, a DBMS could consider all possible candidate plans for a query ({\em exhaustive enumeration})
regardless of the time and space that this requires\footnote{The time and space required by an optimizer will always impact ad hoc queries however.}, and provided it was armed with an accurate cardinality and cost model, would always choose an optimal plan.  In practice,
exhaustive enumeration is infeasible for complex, join-heavy queries and most optimizers ``time-out'' prior to consideration of
all plans for such queries.

One possible approach to measuring optimizer efficiency is to calculate the average time that an optimizer spends optimizing queries
in the benchmark query suite. However, this metric has two notable deficiencies: 1) optimizer times recorded for DBMSs running on different platforms are incomparable, and 2) this metric only measures time and not memory.  The OptMark benchmark
instead measures efficiency using four metrics that specify the size of the search space processed during the optimization of a query:
 
\begin{enumerate}
\item {\lp :} the number of logical plans enumerated,
\item {\jo :} the number of join orderings enumerated,
\item {\pp :} the number of physical plans costed, and
\item {\pj :} the number of physical join plans costed.
\end{enumerate}


It is clear that each of these metrics is a measure of the size of the search space explored for a given query.  But as we show in Table~\ref{t:r2results}, these metrics are also strongly correlated with optimization time.  This table shows the degree of correlation between each metric and the time spent by four different DBMSs optimizing averaged over the 93 join queries from the TPC-DS benchmark.  Correlation is demonstrated with $r^2$ values that show the goodness-of-fit of the linear regression, and that fall between 0.0 and 1.0 with higher values indicating higher correlation.  Note that the most highly correlated metric (shown for each DBMS in boldface) varies from system to system, demonstrating that there is no single ``best'' metric for all systems. However, all metrics have very strong correlations (over 0.7 in all cases) with the optimization time and therefore can be used as predictors of an optimizer's optimization time. An interesting observation is that the $r^2$ value of $\pp$ and $\pj$, and that of $\lp$ and $\jo$ are very close to, or even identical to each other on all the DBMSs. The reason is that non-join plans (e.g., table scan plans and index scan plans) are typically much fewer than join plans, and thus take much less optimization time\cite{timeestimate-sigmod2003}.

As is described in Section~\ref{s_toolkit}, to benchmark a given DBMS one must be able to extract at least one and as many of the four metrics as possible, so that benchmark efficiency results can be compared to as many other DBMSs as possible.

{\bf Linear Regression Results} To collect the results of Table~\ref{t:r2results}, we processed an optimization structure exposed by System X, modified
the optimizer code to add instrumentation to the open source code of PostgreSQL and parsed trace files of MySQL and System Y.  These
techniques allowed us to see (and therefore count) the physical plans that were costed by each DBMS during optimization, and from these results we were able to determine the values of the other metrics: determining physical join plans by removing all non-join physical plans (e.g., group-by plans and index plans) from the physical plans, determining logical plans by converting each physical operator in the physical plan to its logical equivalent and ignoring duplicates, and determining join orderings on the basis of physical join structures.

\begin{figure*}[t]
\centering
\subfigure[MySQL]{
\includegraphics[totalheight=1.0in, angle=0]{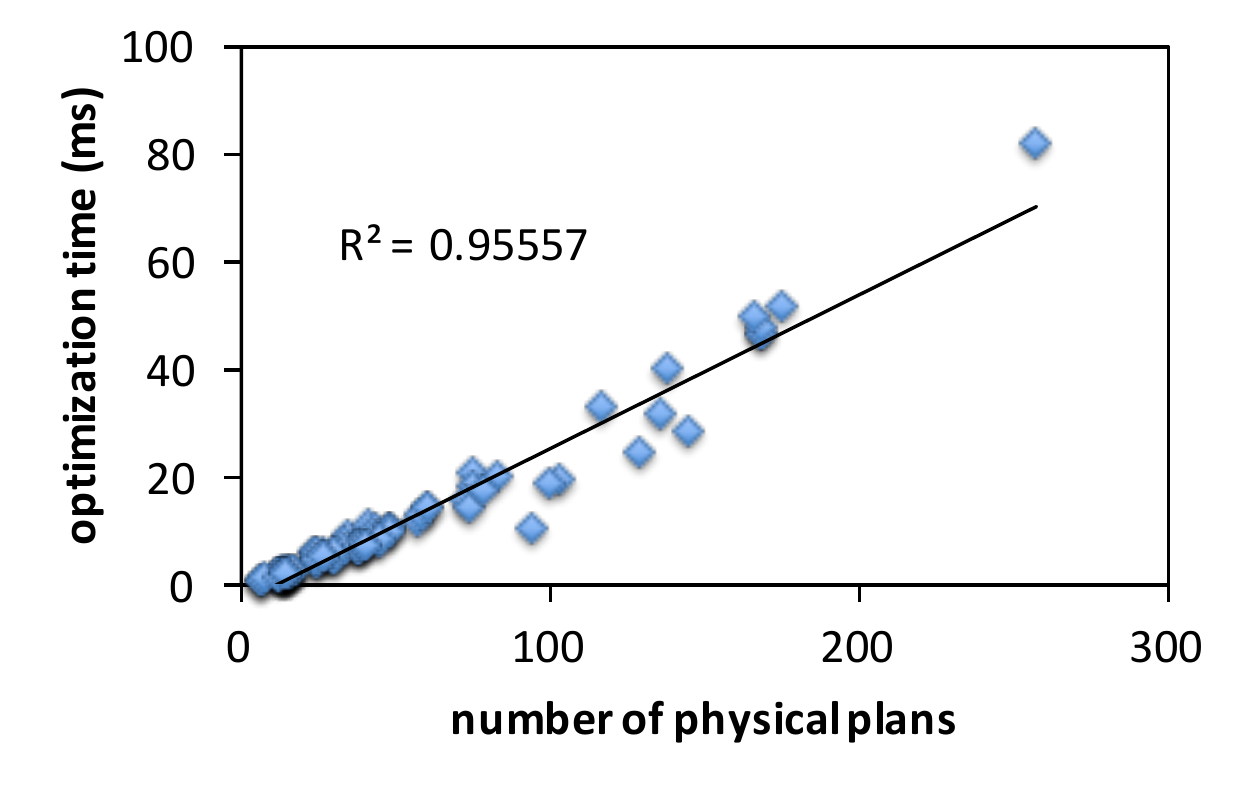}
\label{f:phy-mysql}}
\subfigure[MySQL]{
\includegraphics[totalheight=1.0in, angle=0]{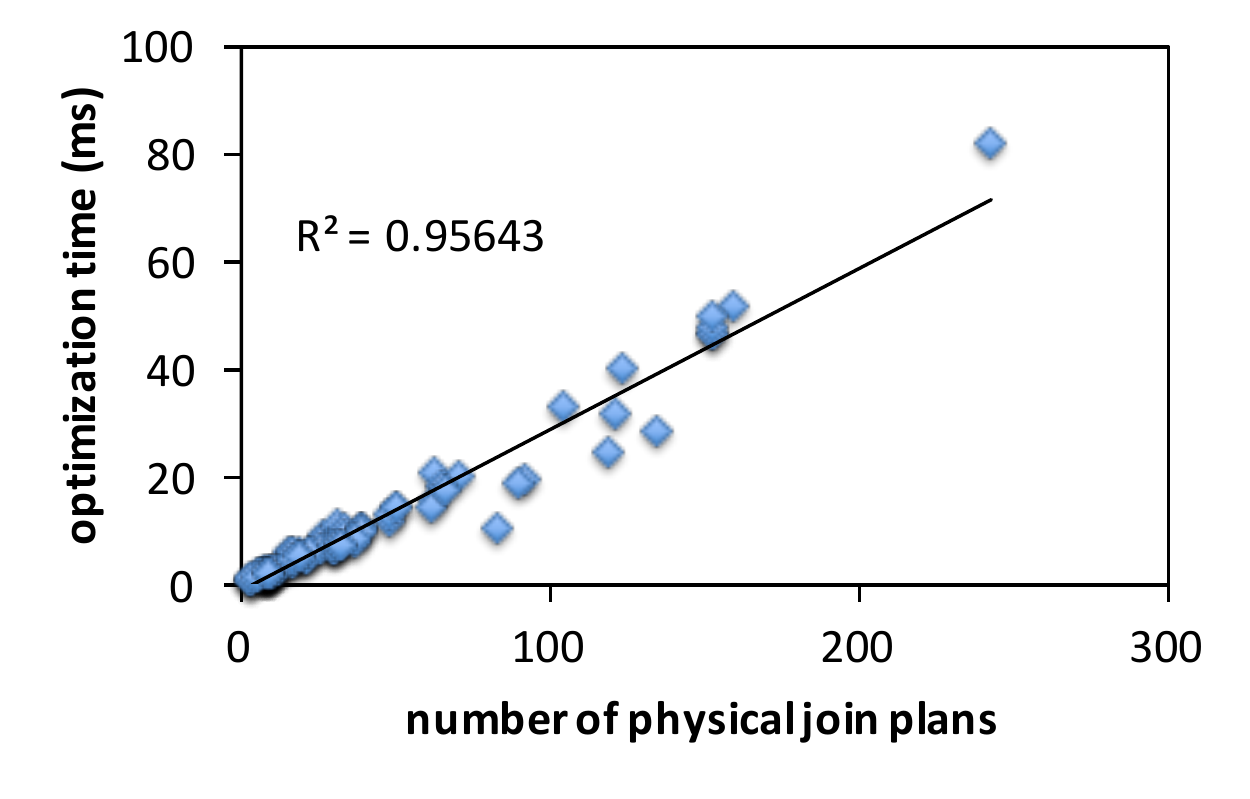}
\label{f:phyjoin-mysql}}
\subfigure[PostgreSQL]{
\includegraphics[totalheight=1.0in, angle=0]{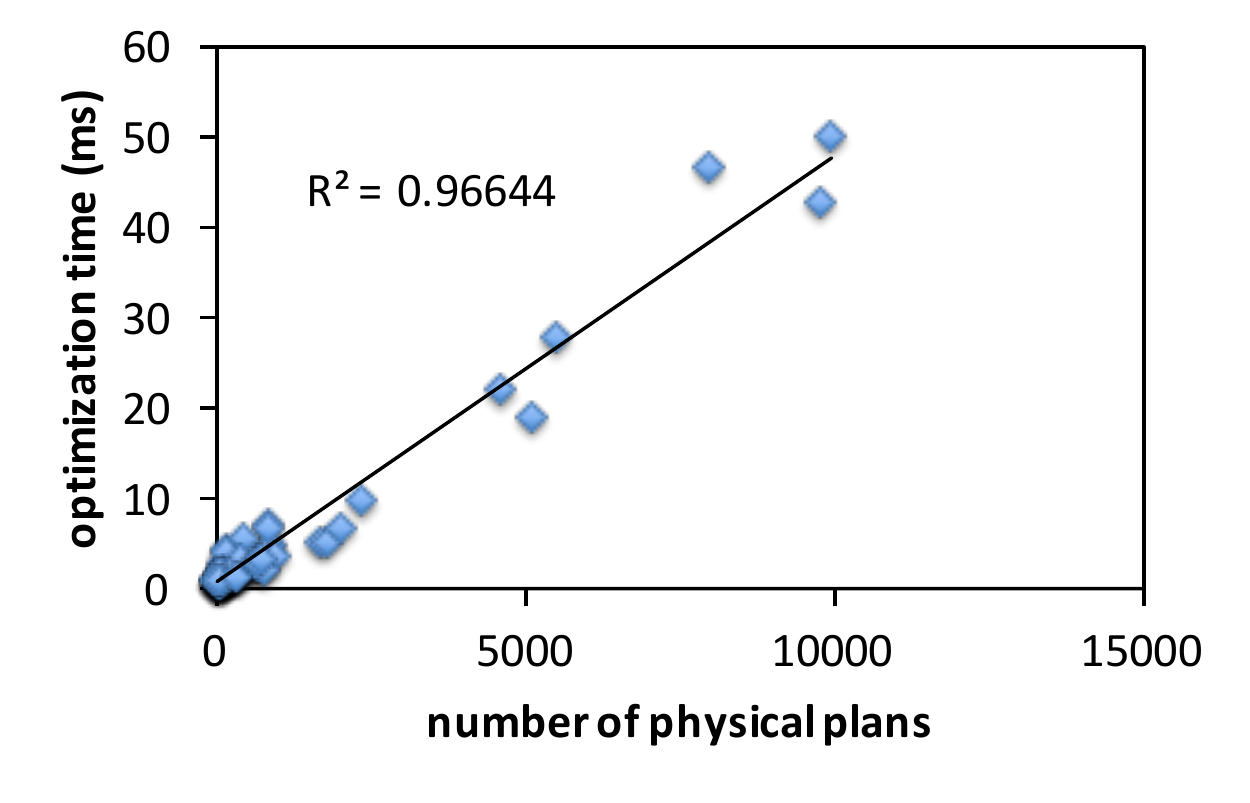}
\label{f:phy-postgres}}
\subfigure[PostgreSQL]{
\includegraphics[totalheight=1.0in, angle=0]{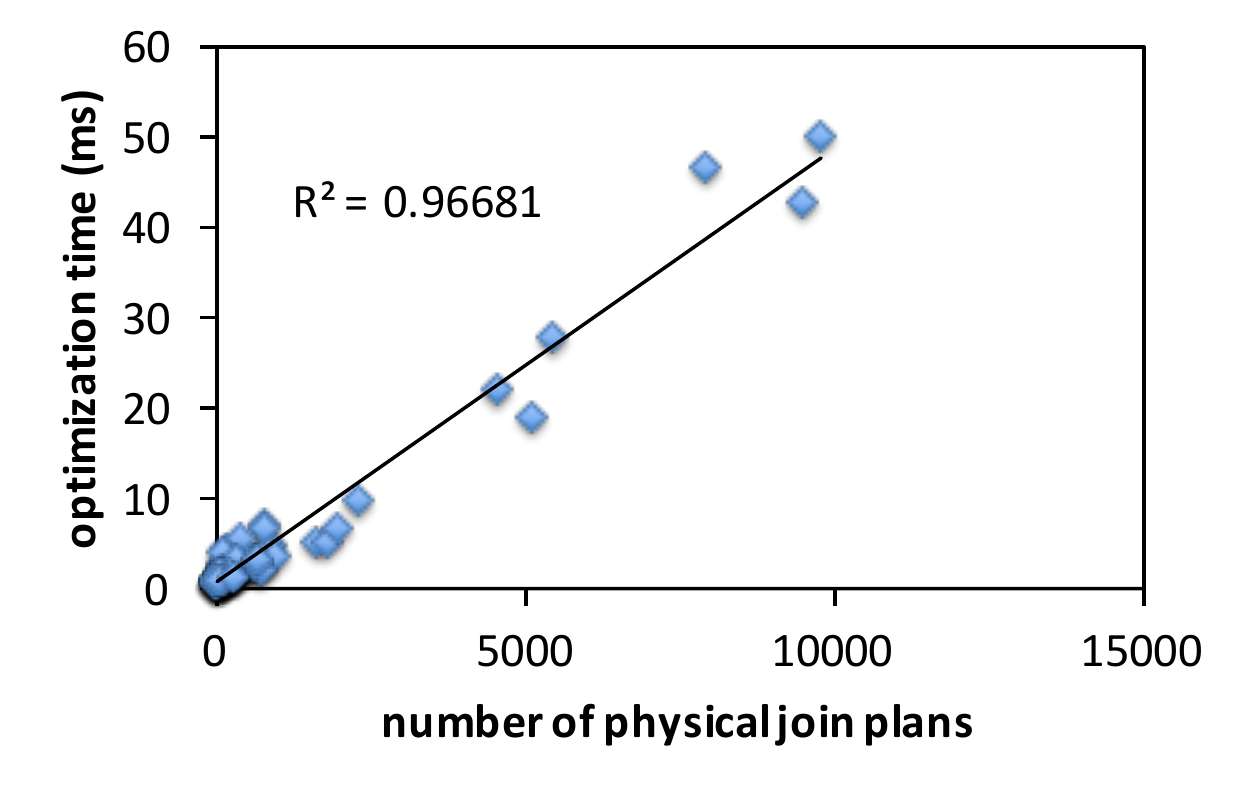}
\label{f:phyjoin-postgres}}
\subfigure[System X]{
\includegraphics[totalheight=1.0in, angle=0]{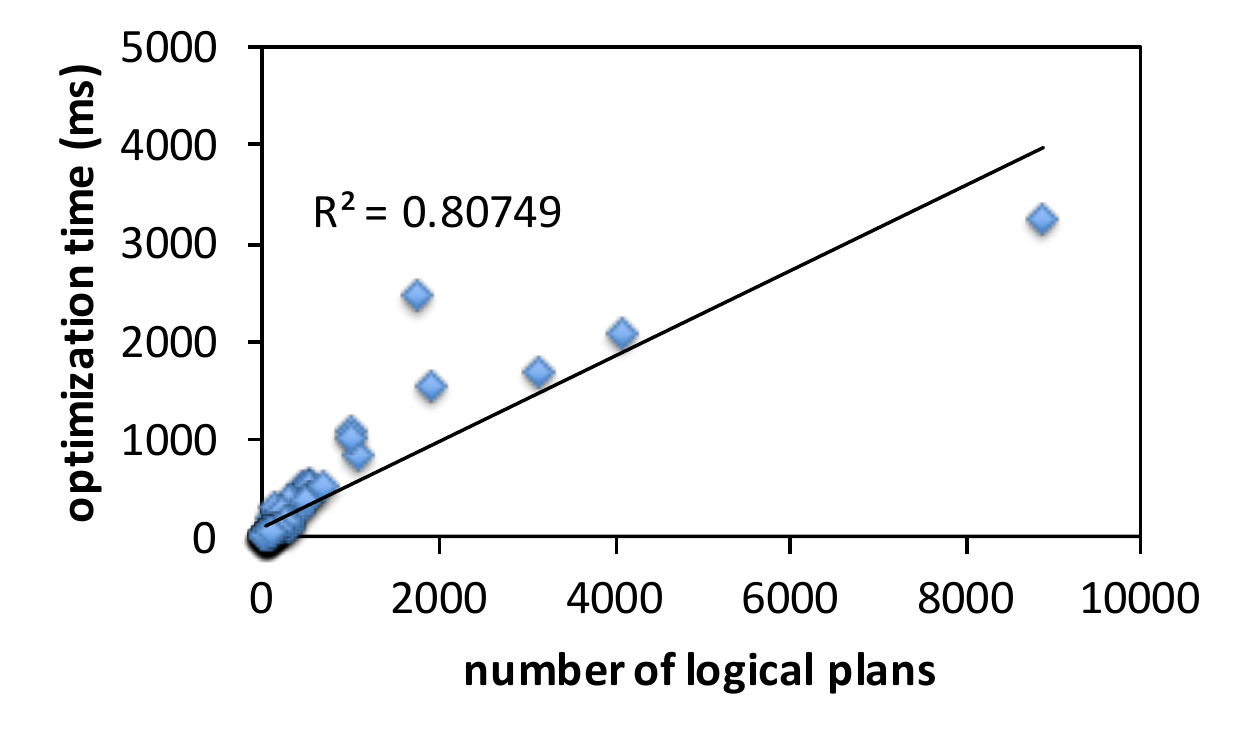}
\label{f:log-sqlserver}}
\subfigure[System X]{
\includegraphics[totalheight=1.0in, angle=0]{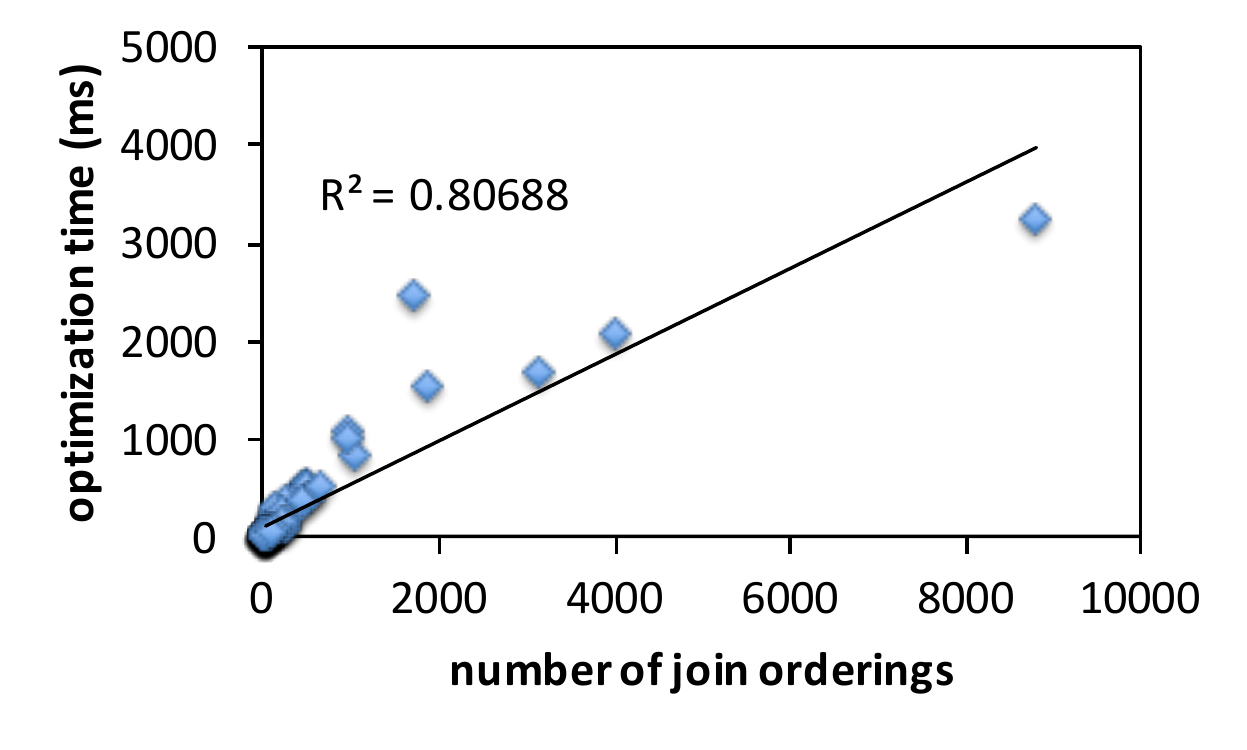}
\label{f:logjoin-sqlserver}}
\subfigure[System Y]{
\includegraphics[totalheight=1.0in, angle=0]{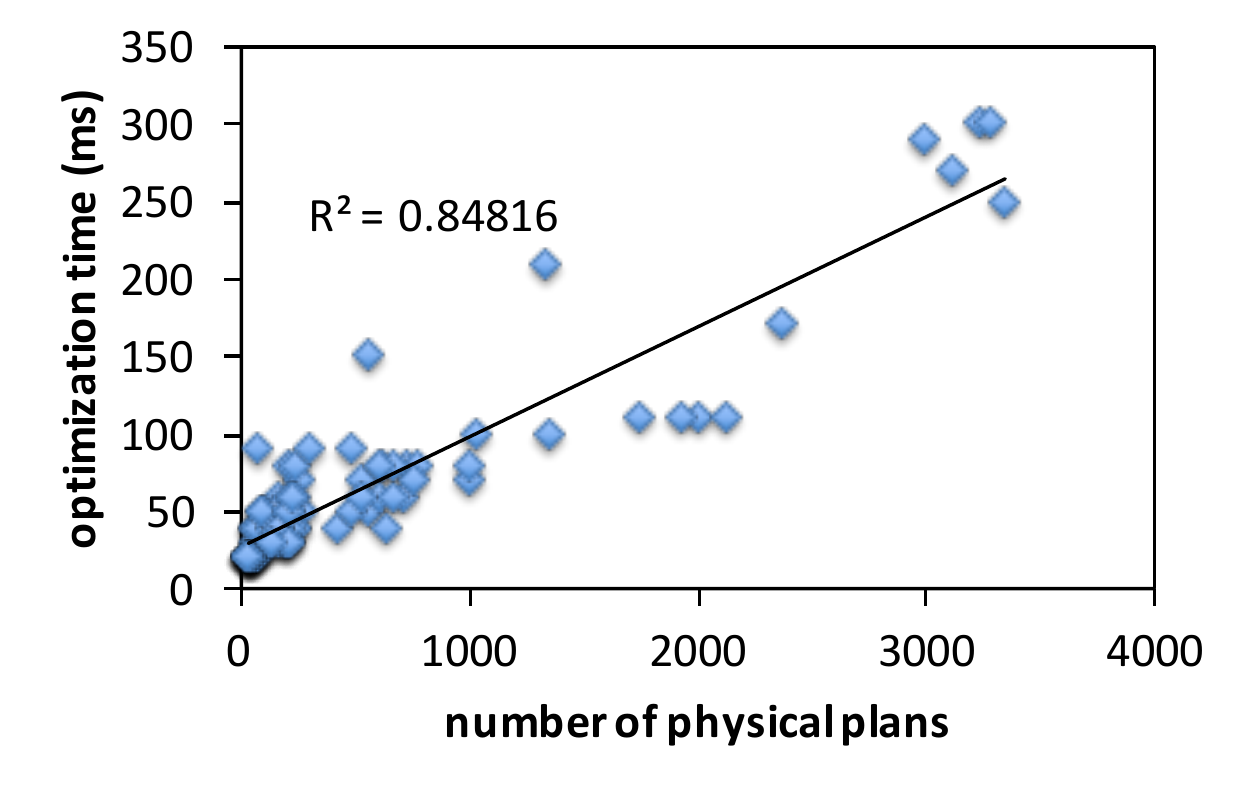}
\label{f:phy-oracle}}
\subfigure[System Y]{
\includegraphics[totalheight=1.0in, angle=0]{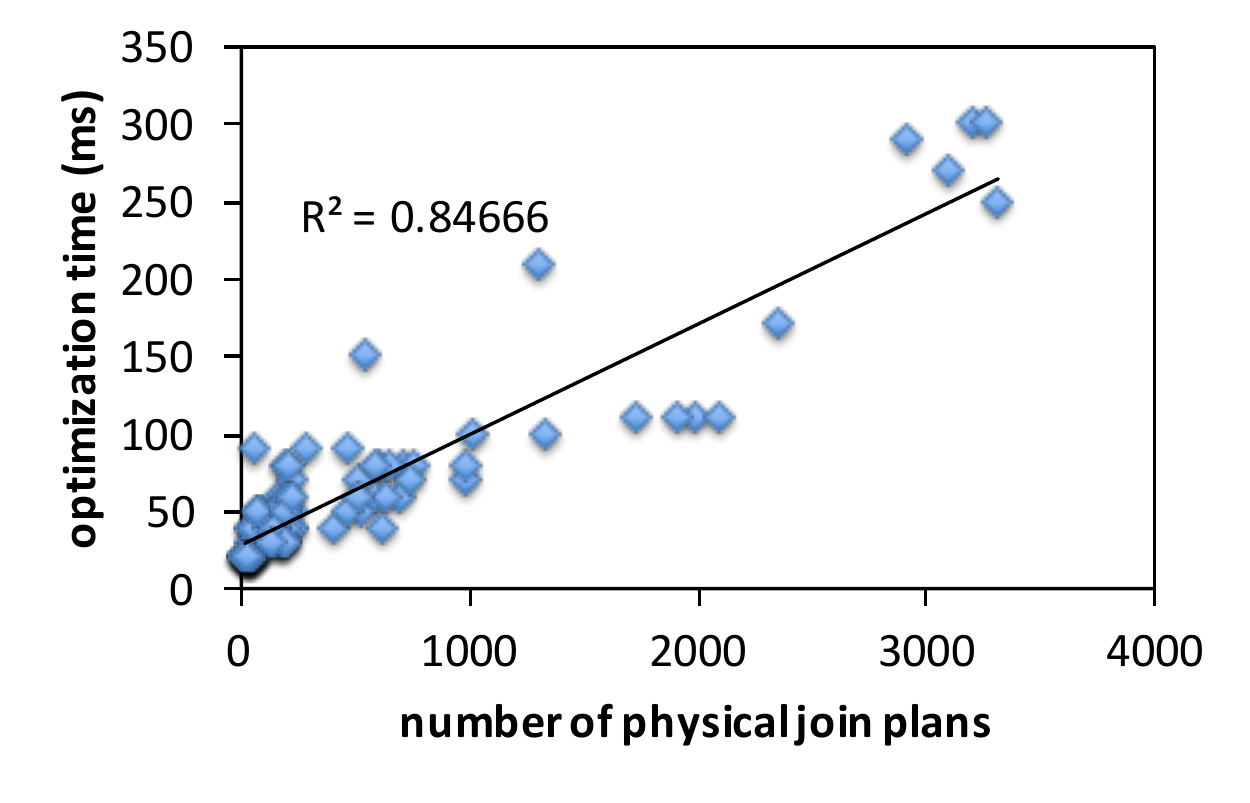}
\label{f:phyjoin-oracle}}
\caption{Linear regression results for four DBMSs.}
\label{f:linear}
\end{figure*}

The regression results demonstrate that all four metrics are good predictors of optimization times as they all offer high $r^2$ values for all four systems. We show the correlation for these metrics for MySQL in Figure~\ref{f:phy-mysql}-\ref{f:phyjoin-mysql}, PostegreSQL in Figure~\ref{f:phy-postgres}-\ref{f:phyjoin-postgres} and System Y in Figure~\ref{f:phy-oracle}-\ref{f:phyjoin-oracle}. This result is an artifact of our technique for extracting logical plans and join orderings from physical plan sets. However, this process potentially overlooks logical plans that were considered by the optimizer but were pruned before being converted to physical ones. These logical plans contributed to the optimization time but since our results are a subset of the actual set of logical plans the correlation appears to be lower for these engines.

MySQL shows a quite strong correlation of the logical plans: almost as high as the correlation of the physical plans. MySQL resolves all joins to nested-loop joins and when mapping logical plans to physical ones it only needs to convert selection operators to table access methods (e.g., table scan, index scan). Hence, in this engine we are able to reconstruct almost the majority of the logical plans, as very few logical ones are pruned before converting to physical ones.  On the other hand, $\lp$ and $\jo$ seems to be a stronger predictors of optimization time than the physical plan metrics $\pp$ and $\pj$ for System X(Figure~\ref{f:log-sqlserver}-\ref{f:logjoin-sqlserver}). Firstly, for this engine we were able to collect \emph{all} logical plans and join orderings considered by the optimizer and hence the correlation is high. However, the data structure offered by this specific engine only reports the physical plans that the optimizer considered to be ``promising'' plans. A high percentage of the pruned plans are not reported although they do add an overhead to the optimization process (since the optimizer examined them). Hence our experiments used a reduced set of physical plans and the regression results thus show a lower correlation of $\pp$ compared with $\lp$.

\begin{table}[t]
\centering

\begin{tabular}{ |c|r|r|r|r|}
\hline
\textbf{System} & \lp & \jo &  \pp & \pj \\

 \hline
  MySQL & 0.92 & 0.93 & \textbf{0.94} & \textbf{0.94} \\ \hline
 PostgreSQL &  0.72 & 0.72  & \textbf{0.97} & \textbf{0.97} \\ \hline
 System X   & \textbf{0.81} & \textbf{0.81} & 0.71 & 0.72 \\ \hline 
 System Y & 0.77  & 0.75  & \textbf{0.85} & \textbf{0.85}\\  
\hline
\end{tabular}

\caption{Correlation of efficiency metrics and optimization times over four DBMSs}
  \label{t:r2results}
\end{table}

\section{The OptMark Toolkit}\label{s_toolkit}

This work introduces OptMark, a query optimizer benchmark toolkit that assists developers measure optimizer effectiveness and efficiency. OptMark is minimum invasive to the underlying database engine. It is executed as a stand-alone tool and runs against a given target database using a standard JDBC interface. It could be extended to any schema and query set, but we want to focus on queries that stress the performance of an optimizer. Typically, these are queries with non-trivia number of joins. Hence, we implemented the toolkit with TPC-DS schema and queries with minimum 5 tables with nesting removed.

Any DBMS for which benchmark results are desired can use our toolkit by implementing a small, simple set of API functions. These functions extract all the optimization metadata and statistics we need to evaluate the effectiveness metrics and efficiency predictors  we discussed in the Sections~\ref{s:effectiveness}-\ref{s:efficiency}. The API consists of two parts:

\begin{enumerate}
\item
Effectiveness API: These functions return the capabilities of query runtime engine (e.g., physical join operations supported) and the means of directing the optimizer to employ specific operations and adhere to specific join ordering.
\item
Efficiency API: The functions return the logical and physical (join) plans that DBMS enumerates and costs while generating plans for a given query.

\end{enumerate}

\subsection{Benchmark Requirements}
For a given DBMS to run the benchmark, it must satisfy the following requirements:
\begin{enumerate}
\item It must support JDBC. The benchmark compares the runtime of the optimizer-chosen plan with the sample set of plans it generates, so OptMark must be able to connect to DBMS through JDBC to execute specific query plans.
\item It must support query hints (or directives) to force the optimizer to consider only plans that adhere to a specific join ordering and/or use specific join operations and access methods.
\item It must expose at least one  of the four predictors we use to assess optimizer efficiency:
\begin{enumerate}
\item number of logical plans or subplans enumerated in optimizing a given query
\item number of join orderings enumerated in optimizing a given query
\item number of physical plans or subplans costed in optimizing a given query
\item number of physical join plans or subplans costed in optimizing a given query
\end{enumerate}
\end{enumerate}

We use requirement 2 to generate a sample set of physical plans with which to compare optimizer generated plan to determine optimizer effectiveness, as described in Section~\ref{s:effectiveness}.
We use requirement 3 to quantify time and space used by optimizer in generating a plan for a query to determine optimizer efficiency as described in Section~\ref{s:efficiency}. Note that not every optimizer exposes each of these indicators, but any two optimizers can be compared if they expose a common indicator.

We have implemented OptMark's API on top of three systems:  two well-known commercial DBMS (which we will call System X and System Y) and the open source engine, MySQL~\cite{mysql}. We note that we were not able to implement the effectiveness assessment of the toolkit on PostgreSQL because OptMark relies on query hints to generate different query plans, which is not supported by PostgreSQL.
Next we discuss the effectiveness and efficiency assessment of OptMark and its API.

\subsection{Effectiveness Assessment}

In Section~\ref{s:effectiveness} we discussed our technique for evaluating the effectiveness of an optimizer. A building block of our approach is enumerating a sample set of physical join plans for a given join query and processing them to collect their execution time. This indicates two requirements for our toolkit. First, it must be aware of the physical join algorithms supported by the underlying DBMS it is executed on (e.g., MySQL supports only nested-loop joins, while DBMSs X and Y support also hash joins and merge-sort joins).  Second, OptMark must be able to enforce the execution of a query using a given physical plan. This plan will specify the join ordering, the physical join operators to be used as well as table access methods. The latter could be either an index-based scan, if an index is available, or a sequential scan. 


Our approach to execute a given physical plan relies on a standard feature in modern databases: query hints that affect the plan choice by the query optimizer.  As the syntax of query hints varies for different systems, OptMark users need to implement a set of  API functions which return the exact syntax of query hints on the system OptMark is running against. This API allows our toolkit to be independent of the underlying DBMS. Next we describe in detail this API.

\subsubsection{Effectiveness API}\label{ss:effectiveapi}

Our effectiveness API consists of three main functions which return (a) the supported physical join operators, (b) the syntax for hinting the use of a given index and (c) the syntax for hinting a specific join method. Next we provide the formal signatures of these API functions.


\begin{enumerate}

  \item {\tt Set\{String\} joinTypes() } \\
  This method returns a list of physical join methods the system supports in the syntax of join hints.
  
  \item
      {\tt String indexHint(String t, String ind) } \\
  Given the table name, {\tt t}, and an applicable index on it, {\tt ind}, the method returns the hint syntax for forcing an index scan on {\tt t} using the index {\tt ind}.

  \item
      {\tt String joinHint(String t1, String al1, String index1,
                String t2, String al2, String index2,
                String join, String clause)} \\
  Given the name, alias and the indexes of two joining tables, the method
  returns the hint syntax to force a two-way join using the specific indexes on each table. The alias and index  parameters are optional. The above API function can also support nested queries. Specifically, the table parameters {\tt t1} and {\tt t2} could be a base table, or a sub-query. If the table parameter is a subquery then the  alias parameter is used as a reference of that query. The index parameters
  {\tt ind1} and {\tt ind2} are  the output of the method {\tt indexHint()},
  i.e., they are the hint syntax for using an index-based table scan. The parameter {\tt join} is
  the output of the method {\tt joinTypes()}, i.e., it is the syntax for forcing the optimizer to use  a specific physical join  operator for executing the query. The parameter {\tt clause} is a string representation of the join conditions to be used.

  \end{enumerate}
  
OptMark also needs to determine which indexes exist on which tables and over which attributes, this information can be got by a JDBC function {\tt getIndexInfo()}.

\subsubsection{Benchmarking Effectiveness}\label{s:effectivenessAlgo}
 
\begin{figure}
\centering
\includegraphics[width=3.2in,angle=0]{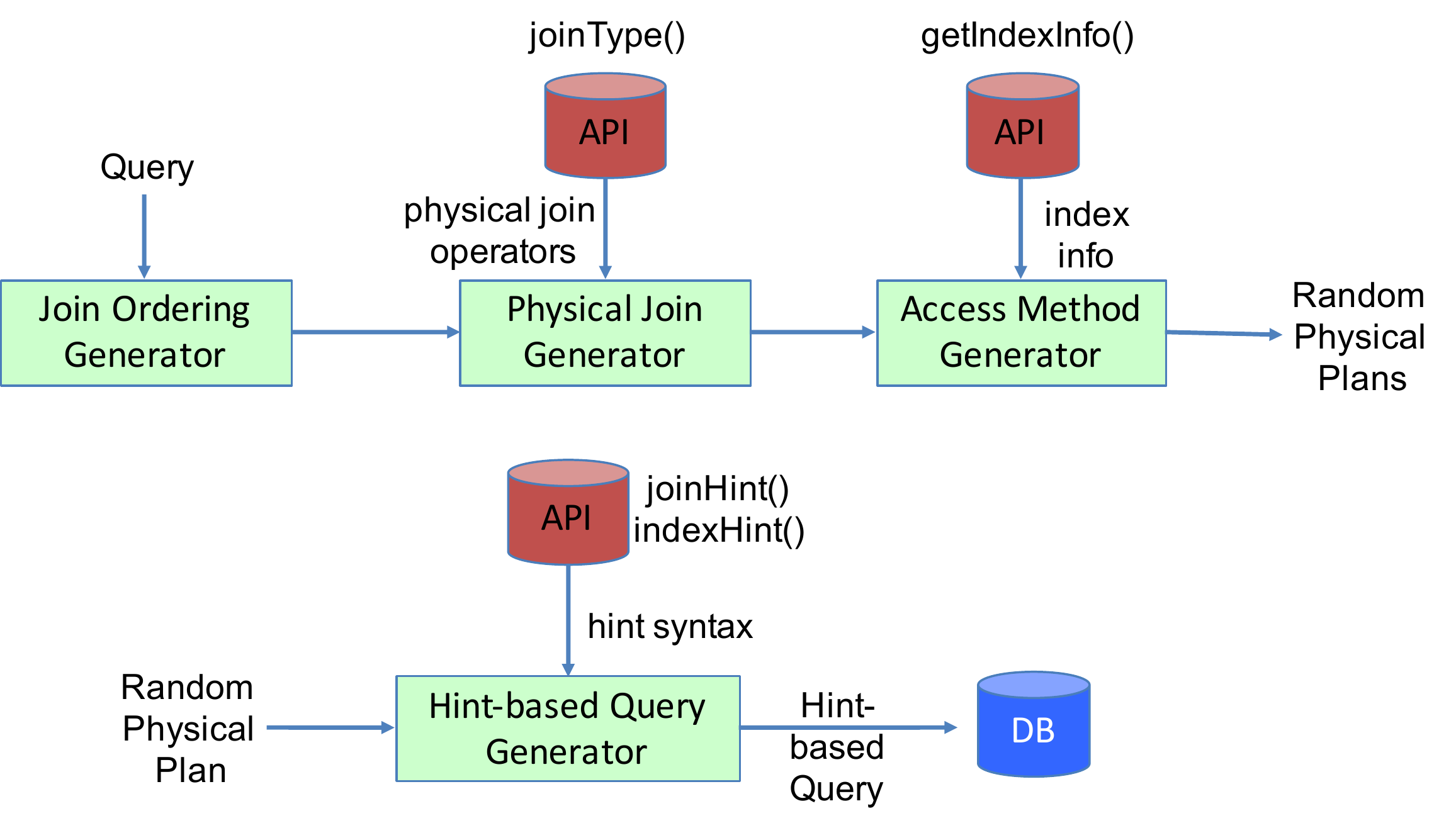}
\caption{Effectiveness Benchmarking Workflow}
\label{f:effectivenessworkflow}
\end{figure}

Next we describe OptMark's approach for assessing the effectiveness of an optimizer. Given a set of queries to execute the benchmark on and an implementation of the effectiveness API for the DBMS OptMark is running on, our toolkit generates a samples set of physical plans against which the optimizer chosen plan is compared with based on the execution time. As discussed in Section~\ref{s:samplesize} the size of the sample set is determined by user-defined confidence level and margin of error. The effectiveness benchmarking workflow for a single query is shown in Figure~\ref{f:effectivenessworkflow}. 

Suppose the given query involves $n$ joins, the random plans with which the optimizer-chosen plan is compared with are generated as follows:

 \begin{enumerate}
 \item First we generate random join ordering as discussed in Section~\ref{s:joinordering}.
 \item For each of the $n$ joins in the join ordering, if there is no join predicate between two tables we force a cross join. Otherwise, we randomly select one physical join operation from all physical join operations returned by {\tt joinType()}. 
 \item For every table in the query we identify if there are any indexes on a predicate attribute (all index information is returned by {\tt getIndexInfo()}) and  we randomly select to use an index or sequential scan to access each table.
 \item Steps 1 - 4 generate a single random  plan. We repeat this process until we collect the sample size determined by confidence level and margin of error the user specified.
 \end{enumerate}

For each plan in our sample set we create a SQL query that enforces the specified join ordering, join algorithm and access method. The  join orderings of step 2 and the choice of join operations in step 3 are enforced according to directives specified in {\tt joinHint()} while the access method is enforced by adding the SQL directive specified by {\tt indexHint()}. The output hint-based SQL query is executed and OptMark collects its runtime and evaluates the query's performance factor. The process is repeated for all queries in a given benchmarking query suite and the optimality frequency of the query set is reported.

\subsection{Efficiency Assessment}

In Section~\ref{s:efficiency} we introduce our four efficiency predictors used by OptMark, namely the number of join and physical plans as well as the number of join orderings and physical join plans. Given a set of benchmark queries, OptMark will report the predictors the DBMS exposes for each of the queries in this query set.

Designing a universal method to extract these predictors which can be used independently of the underlying DBMS was very challenging for the following two main reasons. First, different systems expose the optimization process to a different extent. For example, System X offers a compact representation of the search space of optimization plans which allows us to extract the logical plans. But as it prunes expensive physical plans and only keeps the potentially good ones, we have no way to get  all the actual physical plans the optimizer considers.
	  
Second, DBMSs expose the work of the optimizer in a very different way. For example, System Y and MySQL both provide trace files which records which physical plans the optimizer considered for a given query. But as MySQL resolves all joins to nested-loop joins the trace file only lists the different join orderings the optimizer considers. This differs completely with the trace file of System Y, where it lists all complete physical plans. Though it took us some effort to extract the four optimization time predictors on the four DBMSs we used, we argue that it should be as easy as adding a counter in the optimizer implementation for database developers.

\subsubsection{Efficiency API}\label{ss:efficiencyapi}

 The API that must be implemented in a DBMS specific way to support efficiency assessment of its optimizer should include  at least one of the following functions:
  \begin{enumerate}
  \item {\tt List<PPlan> logicalPlans(Query q)}\\
  Returns all logical plans considered by the optimizer for a given query {\tt q}.
  \item {\tt List<LPlan> logicalJoinPlans(Query q)}\\  Returns all logical join plans considered by the optimizer for a given query {\tt q}.
   \item  {\tt List<LPlan> physicalPlans(Query q)}
   \\ Returns all physical plans considered by the optimizer for a given query {\tt q}.
  \item {\tt List<PPlan> physicalJoinPlans(Query q)}\\  Returns all physical join plans considered by the optimizer for a given query {\tt q}.
  \end{enumerate}
{\tt PPlan} and {\tt LPlan} are physical plans and logical plans in any format the DBMS supports. To assess the efficiency of an optimizer, we count the number of plans returned by any of the above functions  the  DBMS has implemented. For future work, we are looking to reduce the API to return the logical/physical plans and process the results to infer the logical/physical join plans. 

\section{Benchmark Results}\label{s_results}

We have implemented OptMark over three DB systems: MySQL, a commercial system with a top-down optimizer that we refer to as System X,  and a commercial system with a bottom-up optimizer that we refer to as System Y. 

\subsection{Benchmark Environment}\label{ss:setup}

We run our toolkit on a server equipped with a 3.06 GHz Octa CPU and
32 GB of memory. We use the TPC-DS benchmark~\cite{tpcds} for generating our benchmarking dataset and query suite. The workload consists of 24 queries with a  minimum of 5 tables in each query. The benchmark toolkit runs on a dataset of 100GB. 

We have implemented the APIs we described in Section~\ref{ss:effectiveapi} and Section~\ref{ss:efficiencyapi} for System X, System Y and MySQL. To illustrate, we show the System X implementation of the OptMark effectiveness API. The implementation of System Y and MySQL can be found in Appendix.

\begin{enumerate}
  \item 

  {\tt Set\{String\} joinTypes() \{ \\
    ~~return \{HASH, MERGE, LOOP\}; \} } 

 \item 
{\tt String indexHint(String t, String ind) \{ \\
    return WITH (INDEX( + ind + )); \} }

 \item 
{\tt String joinHint(String t1, String al1, String index1,
                String t2, String al2, String index2, 
            String join, String clause)  \{ \\
    return SELECT * FROM  + t1 + index1 + INNER 
    + join + JOIN + t2 + index2 + ON + clause; \} }

  \end{enumerate}

Given the above API, if we want to get the hint syntax from the DBMS for joining two tables A and B using hash join and their corresponding indexes indexA, indexB, one would have to call the function:
\smallskip
  
 \noindent {\tt joinHint(A, A, indexHint(A, indexA), B, B, \\
       indexHint(B, indexB), HASH, A.a = B.b) }, 
\smallskip

\noindent  which would return the hint-based query expression: 
\smallskip

\noindent {\tt SELECT * FROM A WITH (INDEX(indexA)) \\INNER HASH JOIN B WITH (INDEX(indexB)) \\ON A.a = B.b}.
\smallskip

In our discussion we first present the efficiency results then discuss the effectiveness results by three case studies of the three systems on which we implemented OptMark.

\subsection{Efficiency Results} 

We now present the efficiency results. The four metrics, $\lp$, $\jo$, $\pp$, $\pj$, that we use to measure the efficiency of an optimizer were introduced in section 2.2. All metrics have high correlation with optimization time so that any one of them can be used as a metric for efficiency evaluation. The metrics can be used to compare the efficiency of different versions of optimizer. For example, if one introduces a new enumeration strategy, we can use the metrics to check if the new strategy makes the optimizer more efficient. However, one needs to be careful using them to compare different optimizers. For example, the efficiency metrics of MySQL should be less than those of optimizers which supports more physical join operators than just nested loop join which is the only physical join operation MySQL supports. But the comparison between them does not make much sense because they support different join operations. 

As discussed in section~\ref{s_toolkit}, we were not able to implement OptMark on PostgreSQL because PostgreSQL doesn't support query hints. But as PostgreSQL exposes ways to extract physical plans and physical join plans, we can still do efficiency evaluation for PostgreSQL. Table~\ref{t:actualpredictors} shows the values of the four efficiency predictors for the four DBMSs we used. For each predictor, we take the average number over all queries in the workload. Some efficiency predictors are not applicable for some DBMSs because they are not exposed by the DBMSs. From the result we can observe MySQL has the least number of physical plans and physical join plans, which makes sense because MySQL resolves all joins to nested-loop join, its search space is far less smaller than the other systems.

\begin{table}[t]
\centering
\scriptsize
\begin{tabular}{|c|r|r|r|r|}
\hline
\textbf{System} & $\lp$ & $\jo$ &  $\pp$ & $\pj$ \\

 \hline
 MySQL & N/A & N/A & 48 & 40 \\ \hline
 PostgreSQL &  N/A & N/A  & 810 & 778 \\ \hline
 System X & 146 & 123 & N/A & N/A \\ \hline 
 System Y & N/A  & N/A  & 540 & 514 \\  
\hline
\end{tabular}
\caption{Values of efficiency predictors}
  \label{t:actualpredictors}
\end{table}

\subsection{Effectiveness Results}

Next we discuss the  effectiveness results we collected from using OptMark on our  three DBMSs. The assessment process is shown in Figure~\ref{f:effectivenessworkflow}. 
For each of our 24 TPC-DS queries  we generated a random sample set of  385 plans, allowing us to estimate the performance factor of the optimizer with 95\% confidence  and 5\% margin of error. 
 In our discussion, we report the performance factor for each query, as well as the optimality frequency for the given query set. Since we evaluate this metric over a sample set of plans instead of the whole plan space, the results serve as an upper bound of the true optimality frequency.

\begin{table*}
   \centering
    \begin{adjustbox}{width=1\textwidth}
    \begin{tabular}{|c|c|c|c|c|c|c|c|c|c|c|c|c|c|c|c|c|c|c|c|c|c|c|c|c|} \hline
        query&q6&q7&q13&q17&q18&q19&q24&q25&q26&q27&q29&q40&q46&q48&q50&q54&q61&q62&q66&q68&q73&q79&q84&q85 \\ \hline
        {\bf PF}&1&1&0.88&0.917&0.854&1&1&1&0.899&0.989&0.825&0.979&0.94&1&0.969&1&1&0.977&0.982&1&1&1&1&0.912\\ \hline
        \end{tabular}      
    \end{adjustbox}
\caption{Performance Factor (PF) for System X}
\label{t:x-performance}
\end{table*}

\begin{table*}
    \centering
     \begin{adjustbox}{width=1\textwidth}
    \begin{tabular}{|c|c|c|c|c|c|c|c|c|c|c|c|c|c|c|c|c|c|c|c|c|c|c|c|c|} \hline
        query&q6&q7&q13&q17&q18&q19&q24&q25&q26&q27&q29&q40&q46&q48&q50&q54&q61&q62&q66&q68&q73&q79&q84&q85 \\ \hline
        {\bf PF}&1&0.866&1&0.812&0.85&0.871&1&0.89&0.858&0.881&1&0.895&1&0.87&1&1&0.821&1&1&0.92&1&0.885&0.867&1 \\ \hline
    \end{tabular}   
    \end{adjustbox}    
\caption{Performance Factor (PF) for System Y}
\label{t:y-performance}
\end{table*}

\begin{table*}
    \centering
    \begin{adjustbox}{width=1\textwidth}
    \begin{tabular}{|c|c|c|c|c|c|c|c|c|c|c|c|c|c|c|c|c|c|c|c|c|c|c|c|c|} \hline
        query&q6&q7&q13&q17&q18&q19&q24&q25&q26&q27&q29&q40&q46&q48&q50&q54&q61&q62&q66&q68&q73&q79&q84&q85 \\ \hline
        {\bf PF}&1&0.911&0.815&0.711&0.734&0.757&0.871&0.712& 0.816&0.843 &1&0.809&0.864&0.91&1&0.81&1&0.85&1&0.856&0.91&0.88&1&0.751  \\ \hline
    \end{tabular}
    \end{adjustbox}   
\caption{Performance Factor (PF) for MySQL}
\label{t:mysql-performance}
\end{table*}

\subsubsection{Effectiveness of System X}

{\bf Effectiveness metrics}  Table~\ref{t:x-performance} shows the performance factor of each query for System X. The optimality frequency of System X is 0.5, and hence one can conclude  that the optimizer of System X chose the relative optimal plan in no more than 50\% of the queries. The average performance factor for the queries that do not find the relative optimal plan is 0.927. Furthermore, we can say with 95\% confidence that System X finds a plan that is better than 80\% of the generated sample plans (PF$\pm$ 5\% > 80\%) for 96\% of the queries.

{\bf Quality of optimizer-chosen plan} For System X at most 50\% of the queries used the best plan (these are the queries in Table~\ref{t:x-performance} with a performance factor of 1). For the remaining of the queries one should interpret the results along with 95\% confidence and 5\% margin of error we used to generate our sample set of plans. For example, the performance factor of query 13 is 0.88, hence,  with 95\% confidence we can say that the optimizer-chosen plan is better than 88\% $\pm$ 5\% plans in the search space.

An interesting observation is that while the optimality frequency (0.5) of System X shows that in half of the queries the chosen plan was not the best, in 96\% of the queries the chosen plan was better than 80\% of the plans in the sample set. Our results indicate that while the chosen plan by the optimizer of System X might not be the best in half of the cases, it is one of the top plans, which is a good enough plan for an optimizer with limited resources.

{\bf Quality of Cost Model and Plan Enumeration} 
For the queries with performance factor less than 1 one can examine the plans that perform better than the optimizer-chosen plan and get some insight of the quality of the cost model and plan enumeration approach. We examined query 29 which has the lowest performance factor and we discovered that there are 71 plans in the sample that did better than the optimizer-chosen plan. Among them there are 54 plans that were considered by the optimizer, while 17 plans were not considered by the optimizer. As discussed in Section 2.2, we were not able to extract all physical plans the optimizer considered for System X, the number of plans considered here serves as a lower bound. This indicates the cost model fails to estimate accurately the cost of at least 54/71 (76\%) of these better plans. The enumeration quality was however high as the optimizer did not consider only at most 17/71 (24\%) of the better plans.  

For query 18 which also has a relatively low performance factor compared with the other queries, we discovered 56 plans that did better than optimizer-chosen plan. Out of this set, 24 plans were considered by the optimizer, while 32 plans were not. So the cost model fails to accurately estimate the cost of at least 43\% of plans while the enumeration approach did not even consider 57\% of these better plans. 

Finally, to evaluate the efficiency of the optimizer for these queries we collected the number of logical plans. For query 29, the optimizer considers 93 logical plans, while for query 18, the optimizer considers 74 logical plans, indicating the optimizer was more efficient in coming up with a plan for query 18.

\subsubsection{Effectiveness of System Y}

{\bf Effectiveness metrics}  Table~\ref{t:y-performance} shows the performance factor of each query for System Y. The optimality frequency of System X is 0.45. We  conclude  that the optimizer of System Y chose the relative optimal plan in no more than 45\% of the queries. The average performance factor for the queries that do not find the relative optimal plan is 0.868. Finally, we can conclude with 95\% confidence that System Y finds a plan that is is better than 80\% of the generated sample plans (PF$\pm$ 5\% > 80\%) for 87\% of the queries.

{\bf Quality of optimizer-chosen plan} For System Y at most 45\% of the queries used the best plan. For the remaining of the queries one should also interpret the results with 95\% confidence level and 5\% margin of error. For query 25 with performance factor 0.89, one can conclude with 95\% confidence level that the optimizer-chosen plan is better than 89\% $\pm$ 5\% plans in the search space. For System Y, we observed again that while the optimality frequency (0.45) shows that in more than half of the queries the optimizer-chosen plan was not the best, in 87\% of the queries the chosen plan was better than 80\% of the plans in the sample, indicating the optimizer-chosen plan was a good enough plan for an optimizer with limited resources.

{\bf Quality of Cost Model and Plan Enumeration} Aiming for some insight on the quality of the cost model and plan enumeration for System Y, we examined two queries with performance factor less than 1. For query 17, which has the lowest performance factor, among 72 plans that did better than the optimizer-chosen plan, 50 plans were considered and 22 plans were not considered by the optimizer. This indicates the cost model fails to accurately estimate the cost of 70\% of these better plans while the enumeration approach did not consider 30\% of these better plans.
For query 61, which has the second-lowest performance factor, among the 69 plans that did better than optimizer-chosen plan, 45 (65\%) plans were not costed accurately and hence eliminated and 24 (35\%) plans were not even enumerated by the optimizer. We checked also the efficiency of the optimizer for these queries and we discovered that the optimizer costed 735 physical plans for query 17 and 1005 physical plans for query 61. So the optimizer is more efficient for query 17.


\subsubsection{Effectiveness of MySQL}

{\bf Effectiveness metrics} Table~\ref{t:mysql-performance} shows the performance factor of each query for MySQL. The optimality frequency of MySQL is 0.25, and hence one can conclude that the optimizer of MySQL chose the relative optimal plan in no more than 25\% of the queries. The average performance factor for the queries that do not find the relative optimal plan is 0.82. Furthermore, we can say with 95\% confidence that MySQL finds a plan that is better than 80\% of the generated sample plans (PF$\pm$ 5\% > 80\%) for 58\% of the queries.

{\bf Quality of optimizer-chosen plan} For MySQL at most 25\% of the queries used the best plan and  there are still more than half (58\%) of the queries for which the optimizer-chosen plan is better than 80\% of the sample plans. This indicates again while the chosen plan by the optimizer of MySQL wasn't the best in majority of the cases, in more than half of the cases the chosen plan was one of the top plans.

 {\bf Quality of Cost Model and Plan Enumeration} We examined two queries of MySQL. For query 17, which has the lowest performance factor, among 110 plans in the sample that did better than the optimizer-chosen plan, 89 plans were considered and 21 plans were not considered by the optimizer. So the cost model fails to accurately estimate the cost of 81\% of these better plans. The enumeration quality was relatively high as the optimizer did not consider only 19\% of the better plans. For query 85, among 95 plans in the sample that did better than the optimizer-chosen plan, 78 plans were considered and 17 plans were not considered by the optimizer. This indicates the cost model fails to accurately estimate the cost of 82\% of the better plans and the enumeration approach didn't consider 18\% of the plans. Both cases show most of the loss of better plans are because of the inaccuracy of the cost model. The result is expected because plan enumeration is rather simple for MySQL as it only needs to consider join orderings but not physical join operations. As for the efficiency of the optimizer, there are 235 physical plans costed by the optimizer for query 25 and 176 physical plan costed by the optimizer for query 85. Therefore, the optimizer is more efficient for query 85.

\subsubsection{Impact of query complexity}
\begin{figure}[t]
\centering
\includegraphics[width=3in, angle=0]{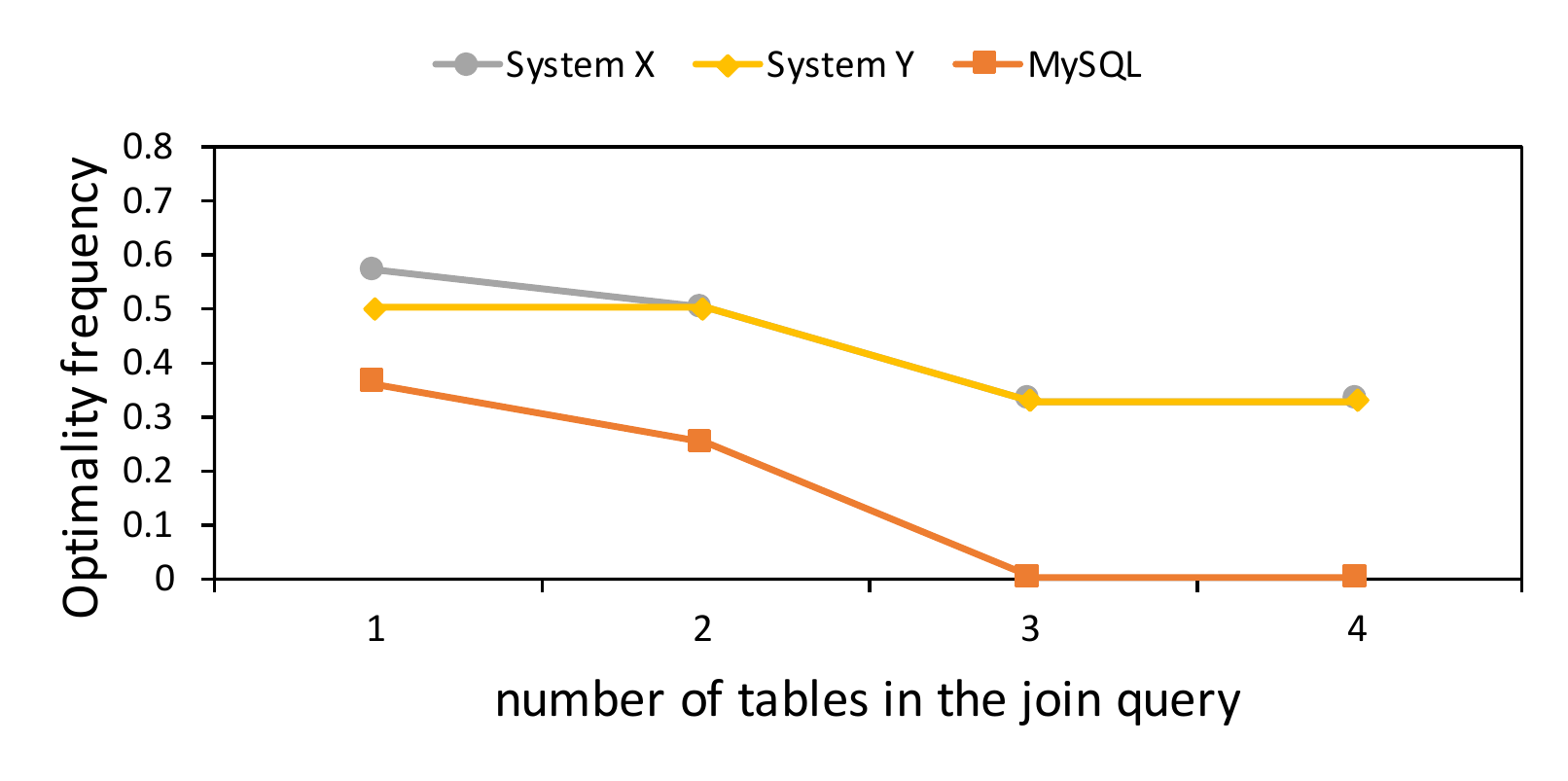}
\caption{Performance factor by query complexity}
\label{f:performance-table}
\end{figure}

In Figure~\ref{f:performance-table}, we report the 
optimality frequency on the three systems as we vary the complexity
of the query. Here, we define the query complexity as the number of tables included in the query. Clearly the more tables involved in a join query the more difficult it becomes for an optimizer to identify the best execution plans. The experiments verified this since  the optimizer optimality frequency decreases  as the number
of tables in our join queries increases. This performance regression is expected because the errors (e.g., cardinality estimation error) the optimizer makes during optimization would ``accumulate" as it needs to estimate join cardinalities for multiple levels of joins. The higher the join level is, the harder it is for the optimizer to choose a good plan. This behavior was consistent on all optimizers.

\section{Related Work}\label{s_related}

Despite of the importance of optimizer benchmarks to date no end-to-end optimizer benchmark is available. \cite{quality-ieee1995} is one of the earliest papers that touch on testing the work of query optimizers and offers a set of tools that support the design and generation of custom testbeds for optimizers. However they do not  provide any measures to evaluate the quality of the produced optimizers. In \cite{challenge-ieee2008} they provide a high-level overview of the unique challenges in testing a query optimizer and possible techniques for validating the optimizer. Our work covers two of the metrics they discussed: optimization time, referred as efficiency in this paper, and query performance which is captured by our effectiveness metrics. A number of papers present tools to assist optimizer benchmarking. \cite{counting-sigmod2000} present algorithms to generate either a whole space of alternative plans, or a uniform random sample. But their research was specific to one system while the way we generate our sample set of plans is applicable to any DBMS. 

There has also been work on testing different components of the query optimizer. \cite{diagrams-vldb2005} introduces a toolkit to visualize the plan space to facilitate the analysis of the cost model and behavior of a plan. In~\cite{space-dbtest11} and \cite{taqo-dbtest12} they focus on the accuracy of cost model. In~\cite{space-dbtest11} they introduce a metric to assess the accuracy across the entire search space while in~\cite{taqo-dbtest12} they develop a framework to quantify the accuracy for a given query workload. The impact of I/O cost estimation on quality of query optimizers has been studied in~\cite{adaptive-icde2007}. In~\cite{howgood-vldb2016} they present ways to quantify the contributions of cardinality estimation, the cost model and the plan enumeration algorithm and provide guidelines for the complete design of a query optimizer.A number of papers addressed the problem of testing cardinality estimation models. \cite{cardestimation-dbtest12} describes the replacement and validation of a new cardinality estimation model in Microsoft SQL Server. \cite{bounding-vldb2009} defines a metric to measure deviations of size estimations from actual sizes. \cite{exactcard-vldb2009} presents a set of techniques that make exact cardinality query optimization a viable option. The effectiveness of transformation rules are studied in \cite{rule-icde2010}.  All of the above work focuses only on testing specific components of the query optimizer, while OptMark provides an end-to-end optimizer benchmark, aiming to reveal overall deficiencies and strengths of the benchmarked optimizer. 

 \cite{timeestimate-sigmod2003} presents a way to estimate the compilation time based on the number of physical plans, which is one of the optimization time predictors we discussed. In contrast with our work, they avoid generating the actual plans and reused the join enumeration process and maintain interesting physical property value to estimate the number of join plans.

As studied in~\cite{howgood-vldb2016}, cardinality estimation errors are usually the main reason for bad plans. Some of our results also reveal this fact. \cite{adaptive-icde2007} test the robustness of an optimizer. 
Finally, there is related work on testing queries instead of query optimizer. \cite{data-vldb2015} addressed the issue of testing SQL queries and automated testing of SQL student assignments. They extend the XData\cite{Shah} data generation techniques to handle a wider variety of SQL queries and a much larger class of mutations.

\section{conclusions}\label{s:conclusions}

This paper introduces OptMark, a toolkit for evaluating the quality of database  optimizers. OptMark offers a set of desirable features to support the assessment of optimizer quality. First, it provides methods for assessing both  effectiveness (i.e., quality of the optimizer's chosen plan) \emph{and} efficiency (i.e., optimization time) of an optimizer. Second OptMark  decouples the evaluation of the optimizer performance from the performance of its underlying DBMSs execution engine, which distinguish it from existing DBMS benchmarks (like TPC). Finally, it is minimum invasive to the underlying engine in that any DBMS for which benchmark results are desired can use our toolkit by implementing a simple set of API functions.

Moving forward we plan to design benchmarks for each component of an optimizer (e.g., cost model, plan enumeration, plan pruning, etc) rather than depending solely on an end-to-end optimizer benchmark. This will facilitate identification of the source of poor plan selection (which could, for example, result from a faulty cost model or plan enumeration strategy that does not enumerate the plans that should be chosen). It will also support the design of robust query optimizer components which are consistent and predictable in the presence of errors in or changes to other components. 

\scriptsize
\bibliographystyle{abbrv}
\bibliography{benchmark,bibfile-mfc}

\clearpage
\newpage
\section*{Appendix}\label{s_appendix}

Here we list optimizer effectiveness API implementation on System X,
System Y and MySQL.

\subsection*{Commercial DBMS: System X}
\begin{enumerate}
  \item 

  {\tt Set\{String\} joinTypes() \{ \\
    return \{HASH, MERGE, LOOP\}; \\
\} } \\

 \item 
{\tt String indexHint(String t, String ind) \{ \\
    return WITH (INDEX( + ind + )); \\
\} }\\

 \item 
{\tt String joinHint(String t1, String al1, String index1,
                String t2, String al2, String index2, 
            String join, String clause)  \{ \\
    return SELECT * FROM  + t1 + index1 + INNER \\
    + join + JOIN + t2 + index2 + ON + clause; \\
\} }\\

  \end{enumerate}

\subsection *{Commercial DBMS: System Y}

\begin{enumerate}
\item

  {\tt Set\{String\} joinTypes() \{ \\
    return \{USE\_HASH, USE\_MERGE, USE\_NL\}; \\
\}} \\

\item
{\tt String indexHint(String t, String ind) \{ \\
    return INDEX( + t + ind + ); \\
\} } \\

\item
{\tt String joinHint(String t1, String al1, String index1,
                String t2, String al2, String index2, 
            String join, String clause) \{ \\
    return SELECT /*+ ORDERED + join + ( + al1 + ,\\
    + al2 + ) + index1 + index2 + */ * FROM + t1 \\ 
    + , + t2 + WHERE + clause; \\
\} }\\

\end{enumerate}

\subsection *{MySQL}

\begin{enumerate}
\item

  {\tt Set\{String\} joinTypes() \{ \\
    return \{LOOP\}; \\
\}} \\

\item
{\tt String indexHint(String t, String ind) \{ \\
    return USE INDEX( + ind + ); \\
\} } \\

\item
{\tt String joinHint(String t1, String al1, String index1,
                String t2, String al2, String index2, 
            String join, String clause)  \{ \\
    return SELECT STRAIGHT\_JOIN * FROM + t1 +  \\
    index1 + , + t2 + index2 + WHERE + clause; \\
\} } \\ 
\end{enumerate}
\end{sloppypar}
\end{document}